\documentclass[aps,prd,nofootinbib,onecolumn,groupedaddress,showpacs,showkeys]{revtex4}

\usepackage{graphicx}
\usepackage{epsfig,latexsym,amssymb}
\usepackage{amsmath}
\usepackage{bm}
\usepackage{slashed}
\usepackage{subfigure}
\usepackage{comment}
\usepackage[dvipsnames]{xcolor}
\usepackage[utf8]{inputenc}
\usepackage[english]{babel}
\usepackage{url}
   \usepackage{multirow}

\newcommand{\Ath}{A_\text{th}}
\newcommand{\Tth}{T_\text{th}}
\newcommand{\fcgc}{f_\text{cgc}}

\newcommand{\be}{\begin{equation}}
\newcommand{\ee}{\end{equation}}
\newcommand{\bea}{\begin{eqnarray}}
\newcommand{\eea}{\end{eqnarray}}

\newcommand{\as}{\alpha_s}

\def\eq#1{{Eq.~(\ref{#1})}}

\newcommand{\ben}{\begin{eqnarray*}}
\newcommand{\een}{\end{eqnarray*}}

\begin{document}


\title{Thermal radiation and inclusive production in the running coupling $k_T$ -- factorization approach}
\author{A. V. Giannini$^{1}$, V.P. Goncalves$^2$ and P.V.R.G. Silva$^2$}
\affiliation{ 
$^1$ Instituto de F\'{i}sica, Universidade de S\~ao Paulo,
Rua do Mat\~ao 1371,  05508-090 S\~ao Paulo-SP, Brazil \\
$^2$ Instituto de F\'{\i}sica e Matem\'atica,  Universidade Federal de Pelotas\\
Caixa Postal 354, CEP 96010-900, Pelotas, RS, Brazil\\
}

\begin{abstract}
The characteristics of the thermal radiation are investigated using a two - component model, with the hard component being described by the Color Glass Condensate formalism. The  inclusive transverse momentum spectra of charged hadrons produced in proton - proton and proton - nucleus collisions at LHC energies and large - $p_T$ are estimated using the running  coupling $k_T$ - factorization formula and the solution of the Balitsky - Kovchegov equation. Our results indicate that the thermal term is necessary to describe the experimental data and that the effective thermal temperature has an energy dependence similar to the saturation scale. We demonstrate that the enhancement of the thermal temperature in $pPb$ collisions is consistent with that predicted by the saturation scale.
\end{abstract}

\maketitle

\section{Introduction}
The description and understanding of the inclusive particle production in hadronic collisions at high energies is still one of the main challenges of the strong interactions theory -- the Quantum Chromodynamics (QCD). While the production of particles with a large transverse momentum $p_T$, usually denoted hard regime, is quite well described by perturbative QCD in terms of the scattering between quarks and gluons, with their subsequent fragmentation, the main contribution for the transverse momentum spectra comes from particles with small $p_T$, which characterizes the soft regime. The mechanism of production of these particles is still not well understood  due to the dominance of nonperturbative effects. In addition, the particle production in the semi -- hard regime, present between the soft and hard regimes, is expected to be modified by nonlinear QCD effects due to high density of gluons in the initial state. At large energies,  the wave function of  the projectiles is probed at small Bjorken $x$, being characterized by a large number of gluons. The dense system of partons is predicted to form a new state of matter - the Color Glass Condensate (CGC) -  where the gluon distribution saturates and nonlinear coherence phenomena dominate (For a review see e.g.~\cite{CGC.review}). Such a system is endowed with a new dynamical momentum scale, the saturation scale $Q_s$, which controls the main characteristic of the particle production and whose evolution  is described by an infinite hierarchy of coupled equations for the correlators of  Wilson lines \cite{BAL,KOVCHEGOV,CGC}. As for large energies $Q_s$  becomes  much larger than the QCD confinement scale $\Lambda_{QCD}$, the presence of the nonlinear effects is expected to affect the soft particle production.

Recent phenomenological studies \cite{Bylinkin:2014vra,Baker:2017wtt,Feal:2018ptp} have demonstrated that the  
transverse momentum distribution of charged hadrons produced 
in high-energy hadronic collisions 
can be accurately described by a two -- component model, given by the sum of an exponential (thermal) component  and a power -- law  (pQCD -- inspired) term. Despite accounting for the bulk of the produced hadrons,
an universal exponential shape for the soft part of 
the spectra is only partially understood. On one hand, such exponential behavior is expected in heavy-ion 
collisions~\cite{Kolb:2003dz} as it is usually associated with the 
thermalization of the produced system, naturally achieved by 
the continuous redistribution of energy through final-state 
interactions. 
On the other hand, its appearance in proton-proton collisions 
is quite surprising. Given the (much lower) final multiplicity 
observed in these collisions, a thermal behavior arising through 
a series of final-state interactions seems quite unlikely. 
Such result has motivated the proposal of alternative models to describe the origin of the thermal behavior in the transverse momenta spectra. 
In Ref.~\cite{Kharzeev:2005iz}, by describing the initial wave function 
of a colliding system according to the CGC formalism, 
the authors associated the thermal behavior to the emergence of 
an event horizon resulting from the rapid deceleration of the colliding 
hadron induced by strong longitudinal color fields, in analogy with 
the Hawking-Unruh effect. The effective temperature of 
the thermal contribution was associated with the hard scale of the 
problem, the saturation scale, $Q_s$. In Refs. ~\cite{Baker:2017wtt,Feal:2018ptp}, the thermal contribution has also been associated to the high degree of quantum entanglement in the hadronic  wave-function. More recently, the connection between the thermal behavior and the CGC formalism has been discussed in Refs. \cite{Gotsman:2019vrv,Gotsman:2019ows}. In particular, such studies have demonstrated that, even changing the model for the hadronization, the transverse momentum spectra cannot be described without a thermal radiation term.


Our goal in this paper is to improve our understanding about the thermal radiation term, mainly focusing on the energy dependence of the effective temperature. As in  Refs. \cite{Bylinkin:2014vra,Baker:2017wtt,Feal:2018ptp}, we will consider the two -- component model, but instead of  a parametrization for the description of the hard component, we will estimate the spectra using the CGC formalism taking into account of the nonlinear effects in the QCD dynamics. In particular, we will use  in our calculations the running coupling $k_T$ -- factorization formula derived in Ref.~\cite{Horowitz:2010yg}  and the solution of the Balitsky -- Kovchegov equation  \cite{BAL,KOVCHEGOV}. As a consequence, we will be able to derive realistic predictions for the spectra in the semi -- hard regime. Such approach  will also allow us to investigate in more detail the magnitude and energy dependence  of the thermal component and the possible relations between the thermal radiation behavior and the CGC formalism. 

The paper is organized as follows. In the next Section, we will present a brief review of the formalism, with particular emphasis in the running coupling $k_T$ -- factorization formula. In Section \ref{sec:res}  the  main parameters of the model are determined using the recent LHC data for the transverse momentum spectra of hadrons in $pp$ and $pPb$ collisions. Results for the energy dependence of the effective temperature are presented and the magnitude of the thermal radiation contribution is estimated. Finally, in Section \ref{sec:conc}, we summarize our main conclusions.

\section{Formalism}

During the last decades, a large amount of data on hadron production in proton -- proton, proton -- nucleus and nucleus -- nucleus has been accumulated. The analysis of these data indicate that  the momentum distribution of charged particles present distinct behaviors depending on the $p_T$
region one is interest. In particular, the data in the soft region (low -- $p_T$), can be described in terms of an exponential function in the transverse mass $m_T = \sqrt{m^2 + p_T^2}$ ($m$ is the hadron mass), similar to the Boltzmann spectrum. As pointed out before, the presence of this thermal behavior in collisions of small objects, as $pp$ collisions, is still a theme of intense debate. On the other hand, the experimental data for the inclusive hadron production at large $p_T$ can quite well be described using perturbative QCD, with the transverse momentum spectra being expressed in terms of the parton distributions of the incident hadrons, fragmentation functions and partonic cross sections. Such distinct behaviors have motivated the authors from Refs.~\cite{Bylinkin:2014vra,Baker:2017wtt} to propose the description of the invariant yield as a sum of the thermal and hard components, as follows 
\begin{equation}
\dfrac{dN}{dyd^2p_T} = F_\text{th}(p_T) +  F_\text{hard} (p_T)\,. 
\label{eq:dndpt}
\end{equation}
As in Refs.~\cite{Bylinkin:2014vra,Baker:2017wtt,Feal:2018ptp,Gotsman:2019vrv,Gotsman:2019ows} we will parametrize the thermal component  by
\begin{equation}
F_\text{th} = A_\text{th}\exp(-m_T/T_\text{th})\,,
\label{eq:thermal}
\end{equation}
where the normalization $\Ath$ and the effective thermal temperature $\Tth$ are free parameters. From  perturbative QCD, it is expected that the hard component will have a $1/p_T^n$ behavior at large transverse momentum, which have motivated the authors to parametrize this component as follows 
\begin{equation}
F_\text{hard} =  A_\text{hard} \left(1 + \frac{m_T^2}{nT_\text{h}^2}\right)^{-n},
\label{eq:thermal_pl}
\end{equation}
where $A_\text{hard}$, $T_\text{h}$ and $n$ are free parameters to be determined from the fit to the experimental data. Although this parametrization captures the main theoretical expectations and allow us to establish a relation between  $T_\text{th}$ and $T_\text{h}$, we believe that a more quantitative study can be performed if instead of  a parametrization for $F_\text{hard}$ we estimate this quantity using a formalism based on first principles, as the CGC formalism. Our study is strongly motivated by Refs.~\cite{Kharzeev:2005iz,Gotsman:2019vrv,Gotsman:2019ows} which demonstrated that the temperature $T_\text{th}$ is proportional to the saturation scale $Q_s$,   establishing a strict connection between the thermal and hard components and sheds a light on the origin of the thermal radiation.

At large energies and central rapidities, the hadronic wave functions of the incident particles are characterized  by a large number gluons with  small values of the Bjorken - $x$ variable.  In this regime, we expect that the QCD dynamics will be described by the CGC effective field theory~\cite{CGC.Raju.McLerran,CGC.review}. Due to the large occupation number of gluons at high energies
the CGC framework assumes the colliding system can be treated classically, meaning the dynamics of particle production are determined by the Classical Yang-Mills equations~\cite{Kovner:1995ja,Kovner:1995ts}.
For proton-proton and proton-nucleus collisions such equations can be linearized and solved 
analytically~\cite{Kovchegov:2001sc,Dumitru:2001ux,Blaizot:2004wu}. 
By doing so, one obtain an $k_{T}$-factorized expression for the inclusive gluon 
production cross section whose main inputs are the unintegrated gluon distributions (UGDs)~\cite{Kovchegov:2001sc} of the incident particles, representing the probability of finding a gluon with 
momentum fraction $x$ and transverse momentum $k_{T}$ in a hadron $h_{i}$. 
By being a product of a classical calculation, the expression derived 
in~\cite{Kovchegov:2001sc} is valid for fixed coupling. The inclusion of 
running coupling effects was later considered in~\cite{Horowitz:2010yg} 
by resumming the subset of Feynman diagrams related with such effects on 
top of the leading order expression. The running coupling corrected 
$k_T$-factorization formula reads\footnote{The
notation follows the one from ref~\cite{Horowitz:2010yg}:
${\bm k}$ denotes the transverse momentum of the produced gluon while
${\bm q}$ and ${\bm k}-{\bm q}$ are the ``intrinsic'' transverse momenta
from the gluon distributions.}:
\begin{eqnarray}\label{eq:rcktfact}
\frac{d \sigma^g}{dy \, d^2 k_T  } \, = \,  \frac{2 \, C_F}{\pi^2} \,
\frac{1}{{\bm k}^2} \,  \int d^2q \,
{\overline \phi}_{h_1} ({\bm q}, x_{1})
\, {\overline \phi}_{h_2} ({\bm k} - {\bm q}, x_{2})
\,
\frac{\as \left(
	\Lambda_\text{coll}^2 \, e^{-5/3} \right)}{\as \left( Q^2 \,
	e^{-5/3} \right) \, \as \left( Q^{* \, 2}\, e^{-5/3} \right)} \,\,,
\end{eqnarray}
where $C_{F}=(N_{c}^{2}-1)/2N_{c}$, $x_{1,2}$ 
being the momentum fraction of the projectile and the target, respectively, and $\Lambda_\text{coll}^2$ is
a collinear infrared cutoff.   Moreover,   ${\overline \phi}_{h_{i}} ({\bm k}, x)$  
denotes the UGD for each colliding hadron~\cite{Horowitz:2010yg},
\begin{equation}\label{eq:rc_ktglueA}
{\overline \phi} ({\bm k}, y) =  \frac{C_F}{(2 \pi)^3} \,
\int d^2 b \,d^2 r \, e^{- i {\bm k} \cdot {\bm r}} \ \nabla^2_r \,
\mathcal{N}_{A} ({\bm r}, y,{\bm b})\,,
\end{equation}
and do not contain a factor of $1/\alpha_s(k^2)$ as in 
the fixed coupling case. Such factors now appear 
explicitly in the denominator of \eq{eq:rcktfact} with the appropriate 
scale whose expression can be found in~\cite{Horowitz:2010yg,nos,Dumitru:2018gjm}. 
The quantity $\mathcal{N}_{A} ({\bm r}, y,{\bm b})$ denotes the
forward dipole scattering amplitude in the adjoint
representation at fixed impact parameter $\bm b$. 
Following previous works~\cite{Dumitru:2018gjm, Levin:2010dw,Levin:2010zy,Levin:2011hr,Dumitru:2011wq,Albacete:2012xq}  
an uniform gluon density within the proton has been assumed. As a direct consequence, $\mathcal{N}_{A} ({\bm r}, y,{\bm b}) = \mathcal{N}_{A} ({\bm r}, y) S({\bm b})$, where $S({\bm b})$ is the profile function of the target, and  
the integration over the impact parameter in \eq{eq:rc_ktglueA} can be performed, generating a factor 
proportional to the effective interaction area of the colliding hadrons. In our study, ${\mathcal N}_A ({\bm r}, y)$ will be  given by the solution of the running 
coupling Balitsky-Kovchegov (rcBK) equation~\cite{rcBK}.
While results for bulk observables in heavy-ion collisions may not 
differ much when using UGD sets associated with different initial 
conditions for the rcBK equation~\cite{Dumitru:2018gjm,Dumitru:2018yjs}, 
the results obtained with the McLerran-Venugopalan model as initial 
condition does not provide the best description of the lepton-hadron 
data at HERA energies~\cite{Albacete:2009fh} nor the $p_{T}$-spectra 
in proton-proton collisions~\cite{Albacete:2012xq}. 
Therefore, in our study we will estimate the gluon spectrum  using the ``g1.101" UGD set 
from Ref.~\cite{Albacete:2012xq}.

In  order to estimate the single inclusive hadron distribution, we need to convolute Eq. (\ref{eq:rcktfact}) with the fragmentation function for gluons into charged hadrons
\begin{eqnarray}
\label{eq:yield_pert}
F_\text{hard} (p_T) = \frac{d N^h}{dy \, d^2 p_T } = \frac{K}{\sigma_{inel}} \int \frac{dz}{z^2} \,\, \frac{d \sigma^g}{dy \, d^2 k_T} \times D_g^h(z = \frac{p_T}{k_T},\mu^2) \,\,,
\end{eqnarray}
where $\sigma_{inel}$ is the inelastic cross section and the ${K}$-factor mimics the effect of higher order corrections and, effectively, of other dynamical effects not included in the CGC formulation. Such factor   is treated as a free parameter, which is a recurrent assumption in phenomenological studies.
In our calculations, we will make use of the  KKP fragmentation functions~\cite{Kniehl:2000fe}.
It is worth noting that the convolution also fix the 
collinear infrared cutoff $\Lambda_\text{coll}^2$ as it 
should match the momentum scale $\mu^2$ used in the fragmentation 
model~\cite{Kovchegov:2007vf}. The above expression is strictly valid in the perturbative regime of large transverse momentum, where the modelling of the hadronization by a fragmentation function can be applied. In particular, the KKP parametrization, as many others fragmentation sets, 
has its application limited to semi-hard momenta, 
usually $p_T \ge 1$ GeV. This limits the application of 
\eq{eq:yield_pert} to this same range. In order to extend the CGC predictions for softer momenta, we will assume that for $p_T < p_{T0} = 1$ GeV, the inclusive hadron distribution can be modelled by a function $A\exp(-m_T^2/Q_0^2)$, where $A$ and $Q_0$ are fixed by imposing the continuity of the function and of its first derivative. 
This ensures the hard component has a continuum behavior with 
$p_{T}$ and that it becomes a constant for $p_T = 0$. While the extension of the hard component to the soft sector is, of course, model dependent, an exponential behavior is in line with the qualitative expectation that the momentum distribution eventually flats out for $p_{T}\rightarrow 0$ due to the nonlinear QCD effects ~\cite{Kovner:1995ja,Krasnitz:2001qu,Krasnitz:2002mn,Kharzeev:2006zm}.

\begin{figure}[t]
	\centering
	\includegraphics[scale=0.4]{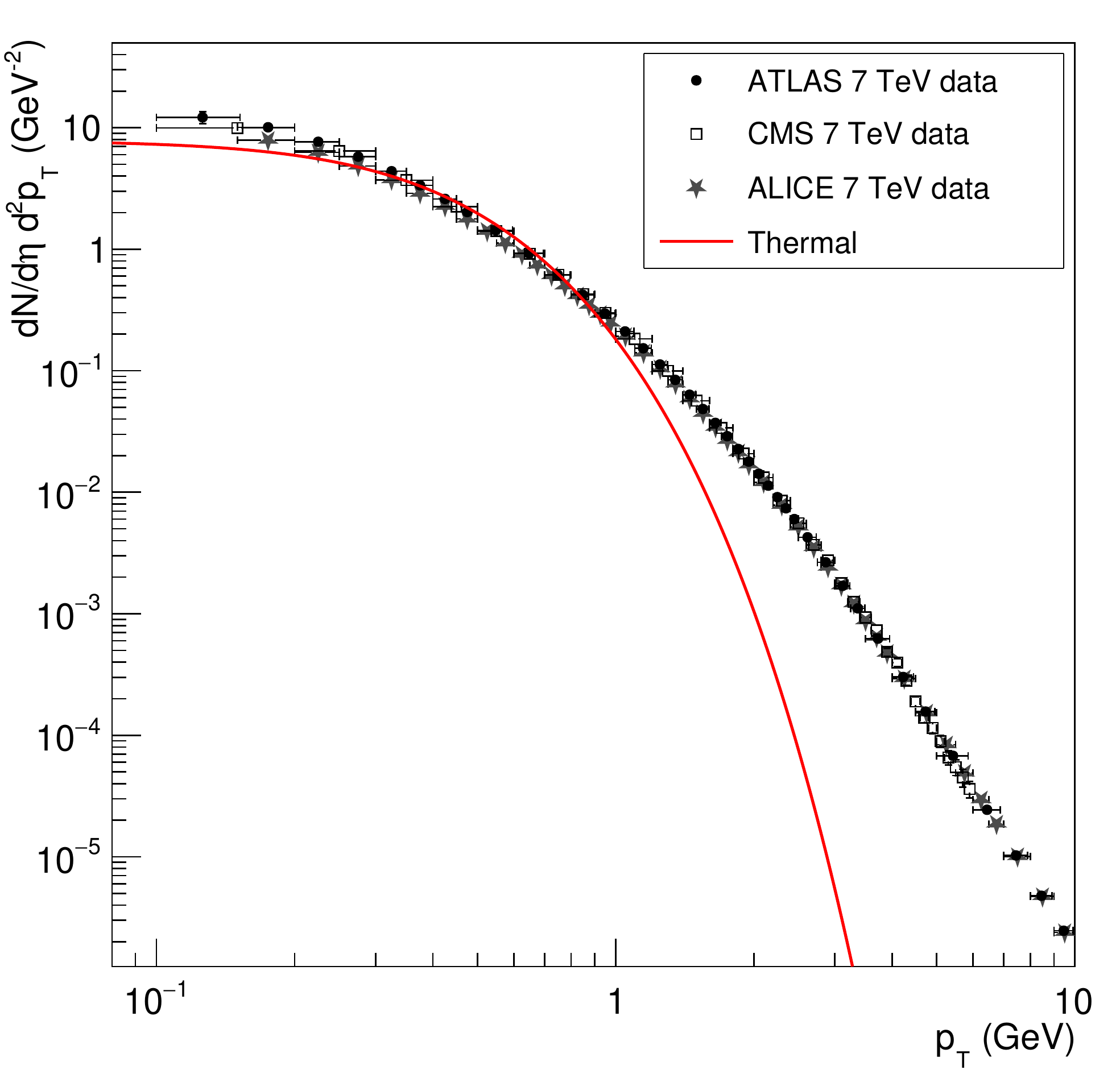}
	\includegraphics[scale=0.4]{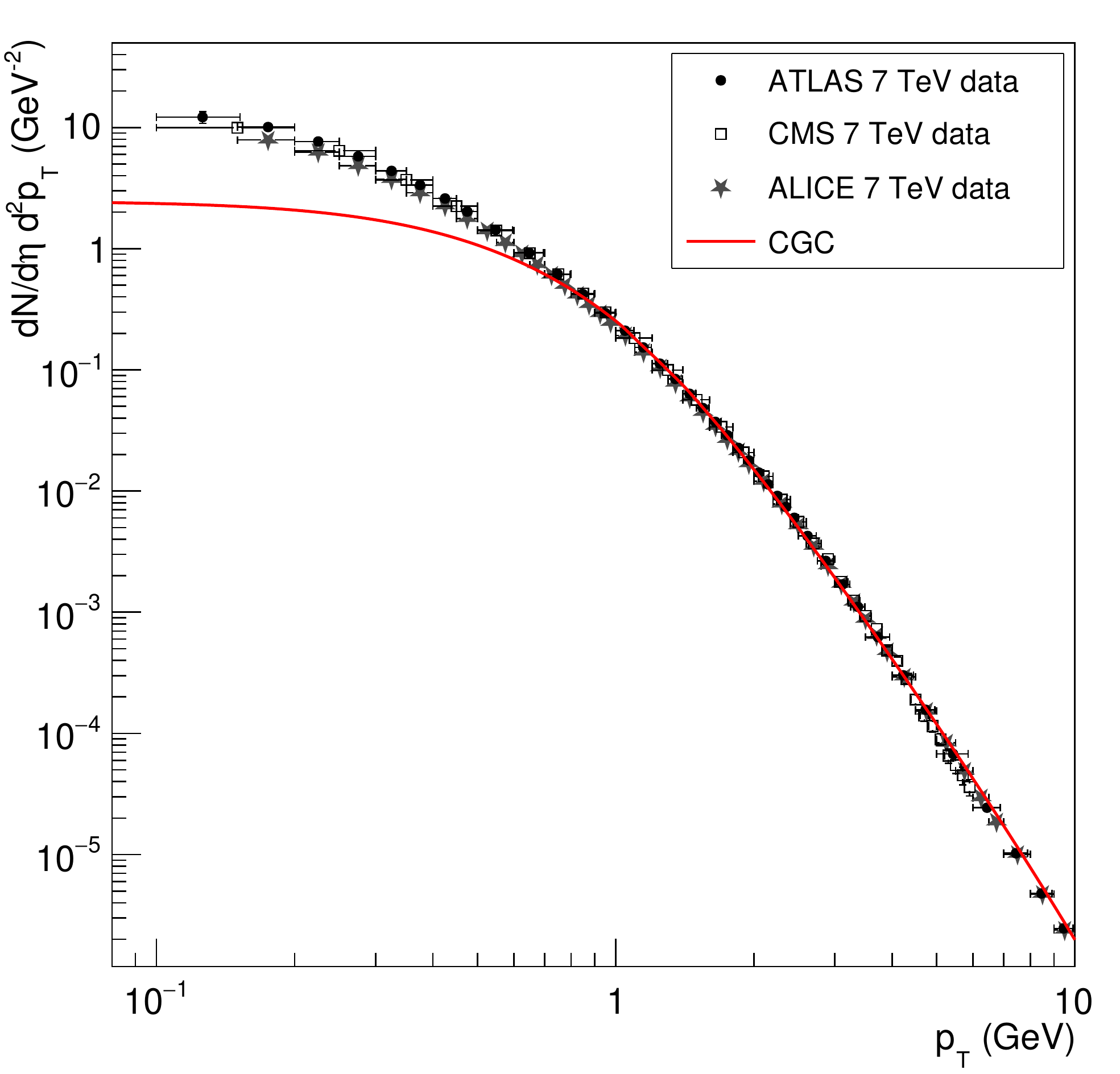}
	\caption{ Results for the inclusive transverse momentum spectra of charged hadrons produced 
		in proton-proton collisions at LHC energy considering only either (left panel) the Thermal or 
		(right panel) the CGC component. Data from ALICE \cite{aliceepjc:2013}, ATLAS \cite{atlasnjp:2010} and CMS \cite{cmsprl:2010} Collaborations at 7 TeV.}
	\label{fig:comp}
\end{figure}

\begin{figure}[t]
 \centering
  \includegraphics[scale=0.75]{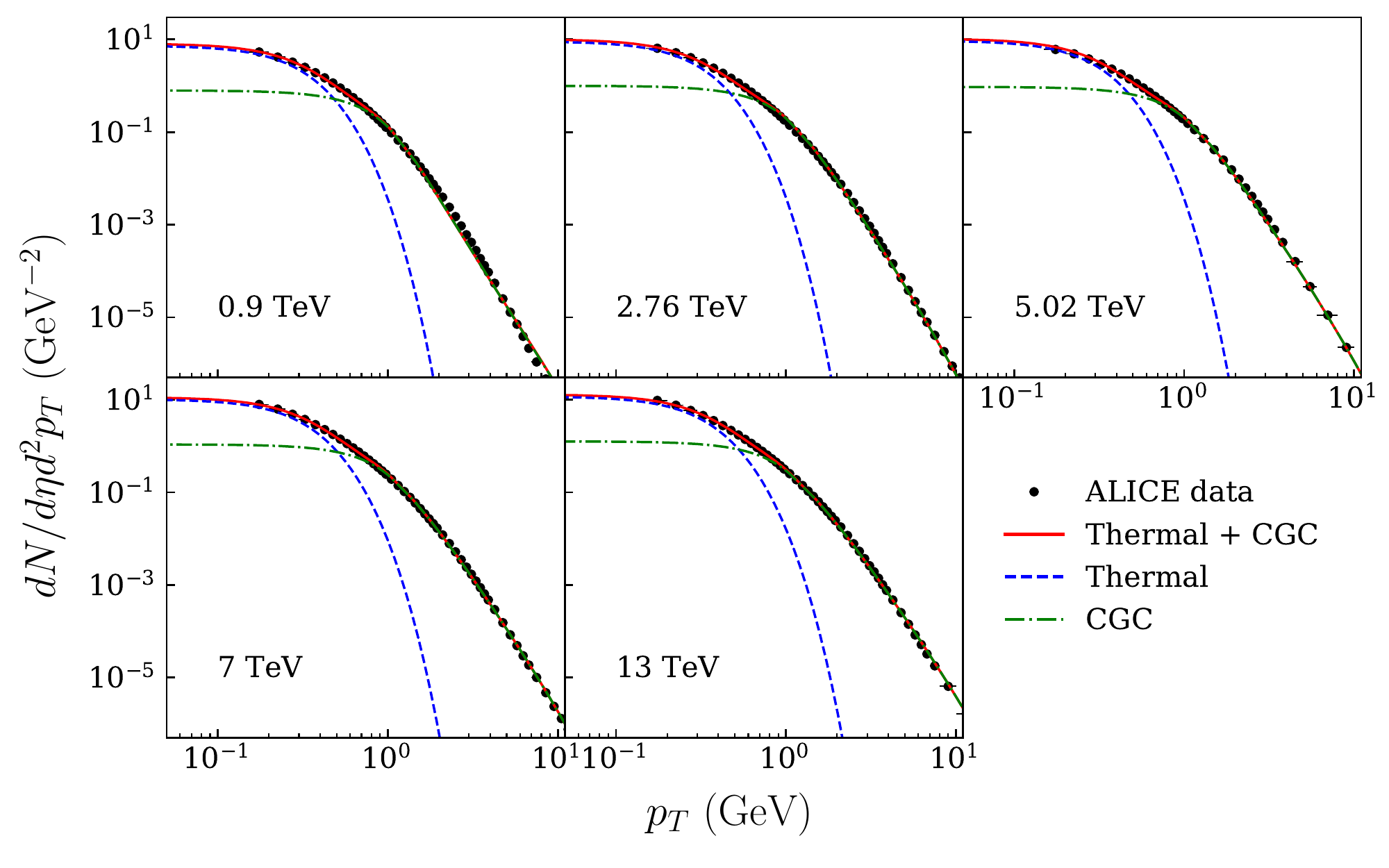}
 \caption{Results for the inclusive transverse momentum spectra of charged hadrons produced 
		in proton-proton collisions for distinct LHC energies considering the  Thermal and CGC components. Data from ALICE Collaboration \cite{aliceepjc:2013,aliceepjc:2013,alicejhep:2018,aliceepjc:2013,aliceplb:2016}.}
 \label{fig:alice}
\end{figure}

\section{Results}
\label{sec:res}
In our analysis of the inclusive hadron production in $pp$ collisions, we will  consider data from different experiments of the LHC in the energy range of 0.9 TeV to 13 TeV. Namely, data by ALICE Collaboration at energies of 0.9~TeV~\cite{aliceepjc:2013}, 2.76~TeV~\cite{aliceepjc:2013}, 5.02~TeV~\cite{alicejhep:2018}, 7~TeV~\cite{aliceepjc:2013} and 13~TeV~\cite{aliceplb:2016}; by ATLAS Collaboration at 7~TeV~\cite{atlasnjp:2010}, 8~TeV~\cite{atlasepjc:2016a} and 13~TeV~\cite{atlasepjc:2016b} and by CMS Collaboration at 0.9~TeV~\cite{cmsjhep:2010}, 2.36~TeV~\cite{cmsjhep:2010} and 7~TeV~\cite{cmsprl:2010}. We included in the fit all data up to $p_{T}^{max} = 10$~GeV, in order to optimize the value of $\chi^2$. Statistic and systematic errors were added in quadrature. Moreover, as experimental data at LHC energies are usually presented in terms of 
pseudorapidity $\eta$ instead of rapidity $y$, the 
transformation $y\to\eta$ in Eq. (\ref{eq:yield_pert}) is accounted for by using the effective 
Jacobian~\cite{Albacete:2012xq} 
\begin{equation}
J(\eta,s) = \dfrac{\cosh\eta}{\sqrt{\cosh^2\eta + [m/P(s)]^2}}\,,
\end{equation}
with $P(s) = 0.13 + 0.32\left(\sqrt{s/s_0}\right)^{0.115}$~GeV 
and  an effective mass of 350 MeV.

\begin{figure}[t]
 \centering
  \includegraphics[scale=0.75]{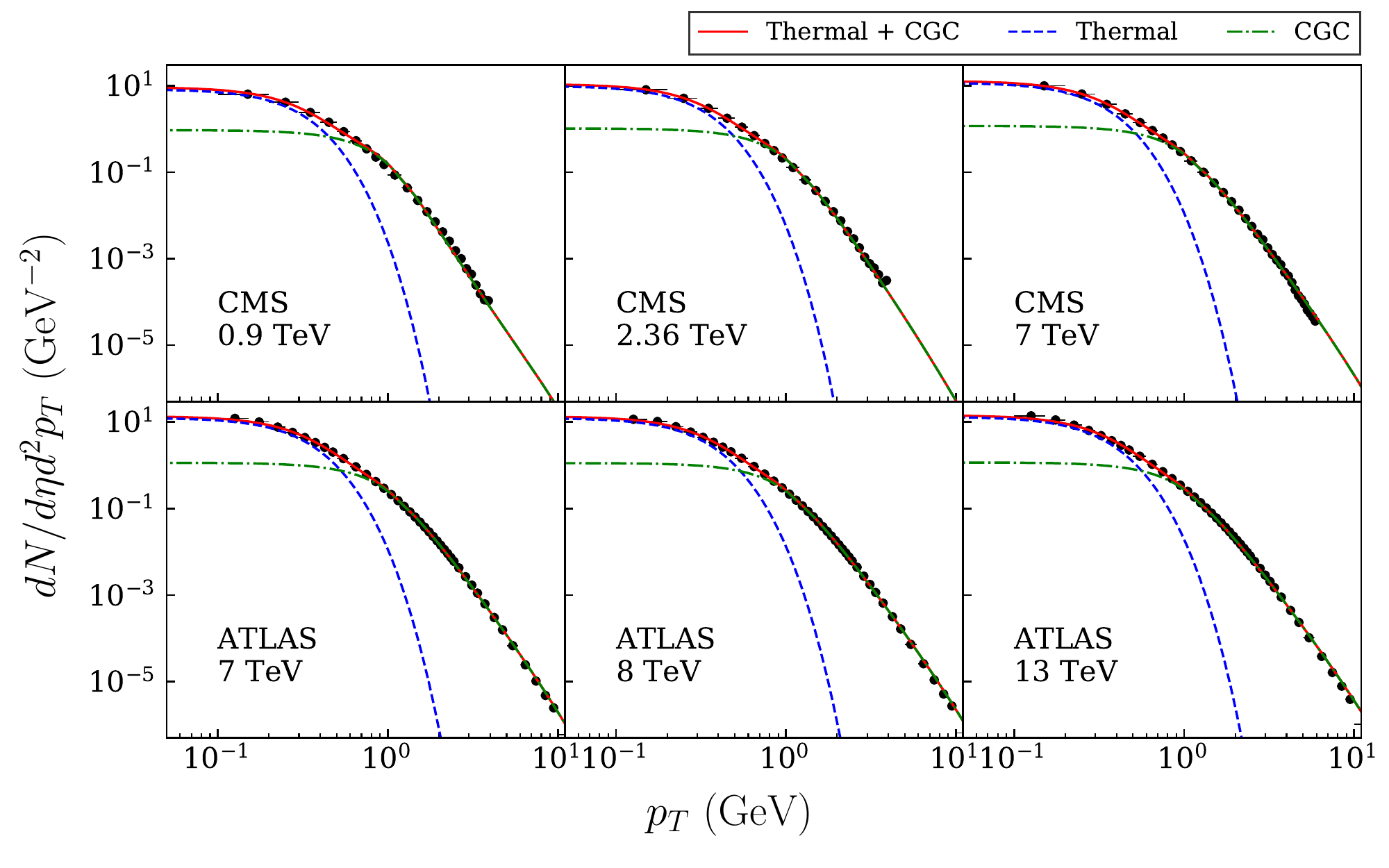}
 \caption{Results for the inclusive transverse momentum spectra of charged hadrons produced 
		in proton-proton collisions for distinct LHC energies considering the  Thermal and CGC components. Data from ATLAS \cite{atlasnjp:2010,atlasepjc:2016a,atlasepjc:2016b} and CMS \cite{cmsjhep:2010,cmsjhep:2010,cmsprl:2010} Collaborations.}
 \label{fig:atlascms}
\end{figure}

Initially, let's investigate the possibility of describe the experimental data for $pp$ collisions at $\sqrt{s} = 7$ TeV considering only one of the terms present in Eq. (\ref{eq:dndpt}). The results of the fit are presented in Fig. \ref{fig:comp} considering either only the soft (Thermal) or the hard (CGC) component. The results indicate that both models are not able to describe the data in the full $p_T$ range considered. In Figs. \ref{fig:alice} and \ref{fig:atlascms} we present our results for the fits of the ALICE and ATLAS/CMS data, respectively, considering the Thermal and CGC contributions. For completeness, the contribution of each component is also shown. As expected, the CGC (Thermal) prediction dominates at large (small) tranverse momentum. It is important to emphasize that the CGC prediction  has only one free parameter, the K-factor, which is energy dependent in agreement with previous studies (See e.g. Refs. \cite{Albacete:2012xq,nospt}). 
A more detailed description of the fits is presented in Table~\ref{tab:results} with some statistical information (reduced $\chi^2$ and degrees of freedom, $\nu$). The parameters without uncertainty are fixed parameters. Such results demonstrate that the two -- component model, Thermal + CGC, describe quite well the current experimental data.

\begin{figure}[t]
 \centering
 \begin{tabular}{cc}
 
  \includegraphics[scale=0.4]{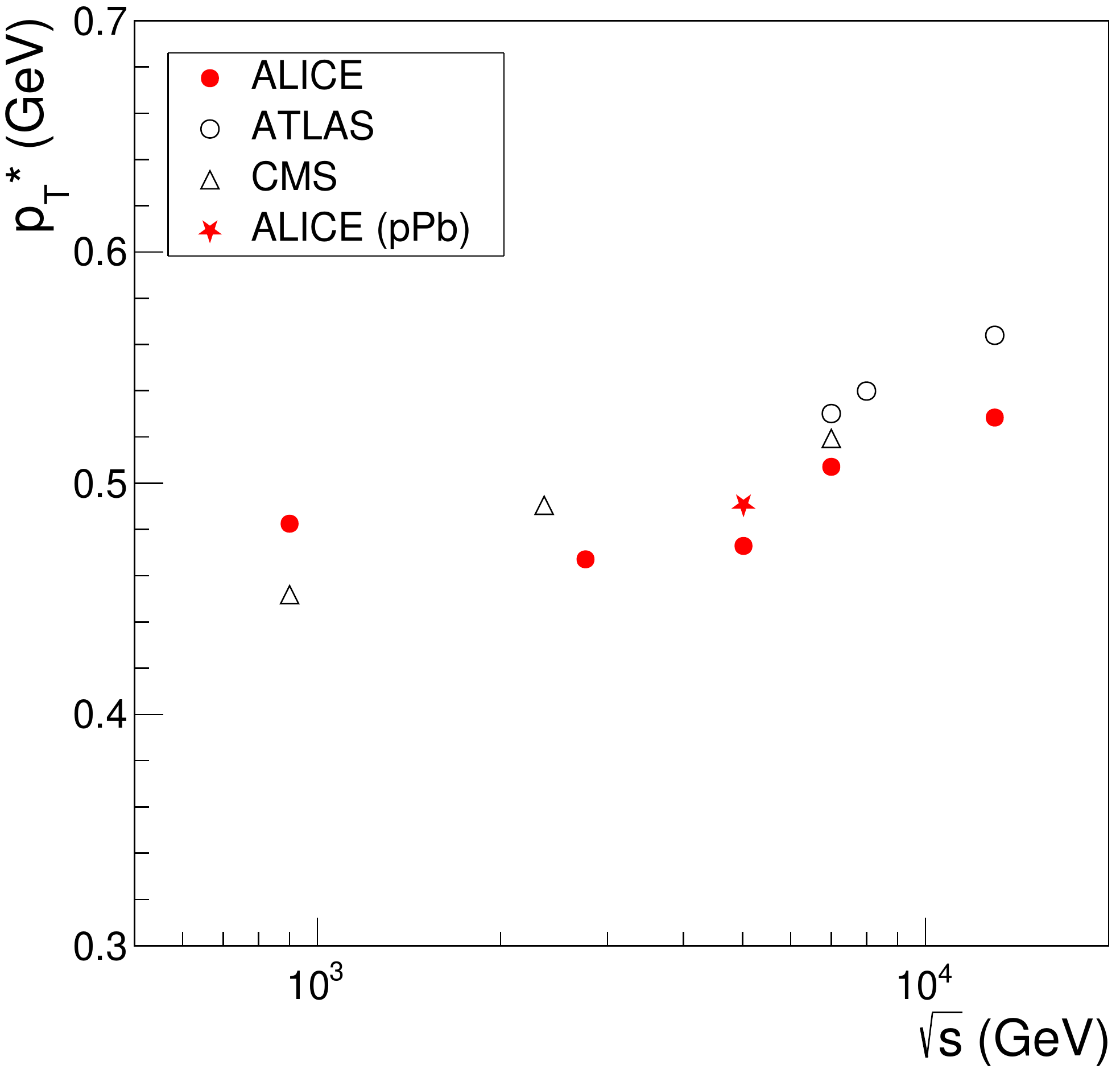}&
   \includegraphics[scale=0.4]{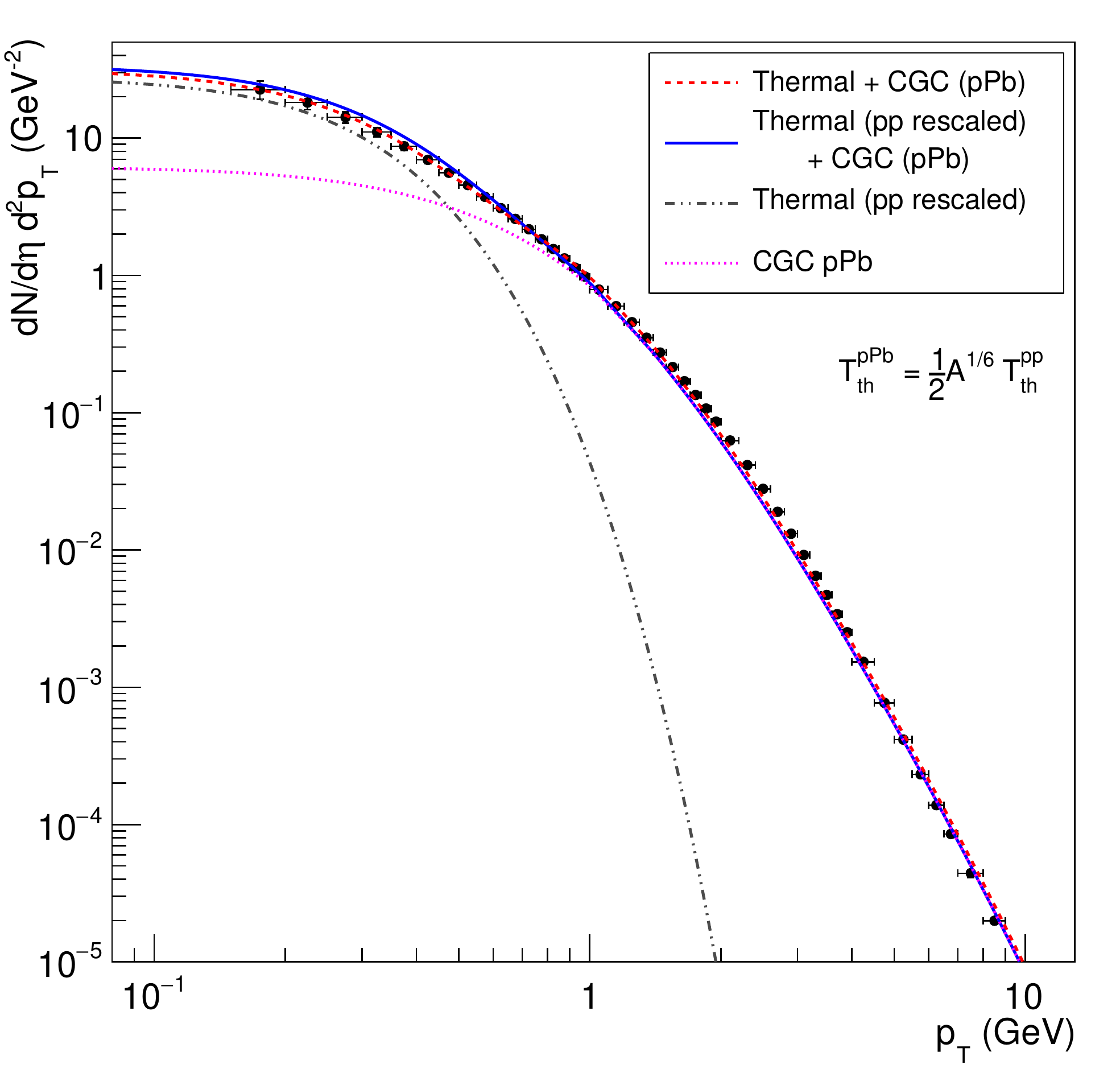} \\
   (a) & (b)
   
 \end{tabular}
 \caption{(a) Energy dependence of the transverse momentum $p_T^*$, where the thermal and the CGC components are equal. (b) Results for the inclusive transverse momentum spectra of charged hadrons produced 
		in $pPb$ collisions at $\sqrt{s} = 5.02$ TeV. Data from ALICE Collaboration \cite{alicejhep:2018}.  }
 \label{fig:ptepPb}
\end{figure}

The results for the fits allow us to estimate the energy dependence of the effective thermal temperature $\Tth$. Assuming that this quantity can be expressed by  
\begin{equation}
 \Tth(s) = T_0 \left[\sqrt{\frac{s}{s_0}}\,\right]^n,
 \label{eq:Tth_energy}
\end{equation}
 \noindent with $s_0 = 1$~GeV$^2$ fixed, we have determined $T_0 = (5.2 \pm 1.3)\times 10^{-2}$~GeV and $n = 0.076 \pm 0.029 $, with $\chi^2/\nu = 0.566$ for $\nu = 9$. One has that the effective temperature increases with the energy and that the value of $n$ is compatible with those derived in Refs. \cite{Bylinkin:2014vra,Gotsman:2019vrv,Gotsman:2019ows}. In particular, it has a similar energy dependence that the saturation scale, in agreement with the theoretical expectation predicted in  Ref.~\cite{Kharzeev:2005iz} and recently confirmed in Refs. \cite{Gotsman:2019vrv,Gotsman:2019ows}.
In order to investigate the impact of the thermal term for different energies, we have estimated the  transverse momentum $p_T^*$, where the thermal and CGC contributions are equal. The results are presented in the left panel of Fig. \ref{fig:ptepPb}. One has that $p_T^*$ slightly increases with the energy, which indicates that the thermal  contribution becomes more important at larger energies. We have verified that for ultrahigh cosmic rays energies, $p_T^*$ is of the order of 0.61 GeV. This result motivates a more detailed analysis of the treatment of hadron production in cosmic rays interactions and possible implications of the thermal radiation for the development of air showers.

During the last years, the ALICE Collaboration has released data for the inclusive transverse momentum spectra of charged hadrons produced in $pPb$ collisions~\cite{alicejhep:2018}. We have estimated the CGC contribution and applied the two -- component model discussed above for these collisions. The results are presented in the right panel of Fig. \ref{fig:ptepPb} by the red dashed line. The associated parameters are the following: $\Ath = 950(313)$~GeV$^{-1}$, $\Tth = 0.1053(64)$~GeV, $K=3.083(46)$, $A = 5.03$~GeV$^{-1}$, $Q_{0}^2 = 0.753$~GeV$^{2}$, with $\chi^2/\nu = 5.46$ for $\nu = 43$.
We have that the model describes the data. Some interesting conclusions can be obtained by the comparison with the $pp$ results. First, the value of $p_T^*$ is slightly larger in $pPb$ than in $pp$ collisions, as shown in the left panel of Fig. \ref{fig:ptepPb}. Such result indicates that the contribution of the thermal term is larger in nuclear collisions, in agreement with the results obtained in Refs.  \cite{Feal:2018ptp,Gotsman:2019ows}.
Second,  Ref.~\cite{Kharzeev:2005iz} has derived that  $\Tth \sim Q_s$. As the nuclear saturation scale is expected to enhanced by a factor $A^{1/3}$ in comparison to the proton one, i.e. $Q^2_{sA} \sim A^{1/3}Q^2_{sp}$, we expect that  ${\Tth^{pPb}}\sim A^{1/6} {\Tth^{pp}}$. In the right panel of 
Fig. \ref{fig:ptepPb} we have tested this expectation, by rescaling the effective thermal temperature obtained for $pp$ collisions at $\sqrt{s} = 5.02$ TeV by the factor $1/2 A^{1/6}$ and estimating the thermal radiation term for this rescaled temperature. The resulting prediction is represented by the blue solid line. One has that the rescaled prediction provides a quite well description of the data, which reinforces the conclusion that the thermal behavior is strictly associated to the treatment of the QCD dynamics at high parton densities, as given by the CGC formalism.

\begin{table}[t]
 \centering
 \begin{tabular}{c||c|c||c||c||c}\hline \hline
  Energy       & \multicolumn{2}{c||}{0.9~TeV} & 2.36~TeV   & 2.76~TeV  & 5.02~TeV   \\\hline
  Experiment   & ALICE            & CMS        &  CMS       & ALICE     & ALICE      \\\hline
  $\Ath$       & 402(134)         & 587(133)   & 485(104)   & 518(182)  & 550(183)   \\      
  $\Tth$       & 0.0930(59)       & 0.0871(38) & 0.0951(41) & 0.0913(62)& 0.0902(51) \\      
  $K$          & 6.27(16)         & 7.47(15)   & 4.54(11)   & 4.040(99) & 2.754(58)  \\      
  $A$          & 1.26             & 1.50       & 1.56       & 1.50      & 1.39       \\      
  $Q_{0}^2$    & 0.551            & 0.551      & 0.608      & 0.616     & 0.646      \\\hline
  $\chi^2/\nu$ & 10.74            & 15.16      & 3.44       & 2.10      & 4.86       \\      
  $\nu$        & 43               & 21         & 21         & 43        & 32         \\\hline\hline
\end{tabular} \\
\vspace*{10pt}
\begin{tabular}{c||c|c|c||c||c|c}\hline\hline
  Energy       & \multicolumn{3}{c||}{7~TeV}         & 8~TeV      & \multicolumn{2}{c}{13~TeV}  \\\hline
  Experiment   & ALICE      & ATLAS      & CMS       & ATLAS      & ALICE      & ATLAS          \\\hline
  $\Ath$       & 389(126)   & 476(89)    & 434(110)  & 435(47)    & 358(70)    & 393(51)        \\
  $\Tth$       & 0.1012(67) & 0.1013(37) & 0.1021(55)& 0.1036(22) & 0.1075(45) & 0.1085(30)     \\
  $K$          & 2.659(68)  & 2.822(33)  & 2.903(71) & 2.608(18)  & 2.284(52)  & 2.125(19)      \\
  $A$          & 1.56       & 1.66       & 1.71      & 1.63       & 1.77       & 1.65           \\
  $Q_{0}^2$    & 0.661      & 0.661      & 0.661     & 0.667      & 0.687      & 0.687          \\\hline
  $\chi^2/\nu$ & 0.739      & 4.14       & 1.99      & 9.49       & 0.992      & 17.2           \\
  $\nu$        & 43         & 37         & 31        & 37         & 42         & 38             \\\hline\hline
\end{tabular}
 \caption{\label{tab:results} Results of fits to experimental data together with statistical information: reduced $\chi^2$ and degrees of freedom, $\nu$. Parameters without uncertainty were fixed in the fits. In all fits we considered 1$\sigma$ of confidence level. Parameter $\Ath$ and $A$ are given in GeV$^{-2}$, $\Tth$ and $T$ in GeV while $K$ is dimensionless.}
\end{table}


\section{Conclusions}
\label{sec:conc}
One of the more intriguing aspects of the description  of inclusive particle production in hadronic collisions at high energies is 
the presence of a thermal radiation behavior at small transverse momentum. For $pp$ collisions, the impact of final state interactions is expected to be small since  such behavior, usually associated to the thermalization of the produced system, is not predicted to be present. Such result have motivated several authors to investigate the connection between the thermal behavior and the high partonic density present in the initial state. At large densities, the nonlinear QCD effects are expected to modify the particle production and to imply the emergence of the thermal spectrum characterized by a temperature proportional to the saturation scale. In this paper we have investigated the characteristics of the thermal radiation term assuming that the hard component of the  two -- component model is described by the running coupling $k_T$ -- factorization formula, calculated using the solution of the Balitsky -- Kovchegov equation. Our analysis improve previous studies that have assumed a parametrization for the hard component. We have demonstrated that the thermal component is needed to describe the inclusive transverse momentum spectra of charged hadrons produced  in $pp$ and $pPb$ collisions at the LHC energies. In particular, we shown that this component becomes more important in nuclear collisions and that the nuclear dependence of the effective thermal temperature is compatible with that expected from the relation with the saturation scale.

\begin{acknowledgments}
This work was  partially financed by the Brazilian funding
agencies CNPq,   FAPERGS and INCT-FNA (process number 
464898/2014-5 and 155628/2018-6).
A.V.G. acknowledges the Brazilian funding agency FAPESP
for financial support through grants 2017/14974-8 and 2018/23677-0.
\end{acknowledgments}

\end{document}